# A Decision Theoretic Perspective on Artificial Superintelligence: Coping with Missing Data Problems in Prediction and Treatment Choice

Jeff Dominitz
Department of Economics
Rice University, Houston, TX 77005, USA
jd232@rice.edu

Charles F. Manski
Department of Economics and Institute for Policy Research
Northwestern University, Evanston, IL 60208, USA
cfmanski@northwestern.edu, ORCID: 0000-0001-7260-7686

## Abstract

Enormous attention and resources are being devoted to the quest for artificial general intelligence and, even more ambitiously, artificial superintelligence. We wonder about the implications for methodological research that aims to help decision makers cope with what econometricians call *identification problems*, inferential problems in empirical research that do not diminish as sample size grows. Of particular concern are missing data problems in prediction and treatment choice. Essentially all data collection intended to inform decision making is subject to missing data, which gives rise to identification problems. Thus far, we see no indication that the current dominant architecture of machine learning (ML)-based artificial intelligence (AI) systems will outperform humans in this context. In this paper, we explain why we have reached this conclusion and why we see the missing data problem as a cautionary case study in the quest for superintelligence more generally. We first discuss the concept of intelligence, focusing initially on some work by AI researchers, before presenting a decision-theoretic perspective that formalizes the connection between intelligence and identification problems. We next apply this perspective to two leading cases of missing data problems. Then we explain why we are skeptical that AI research is currently on a path toward machines doing better than humans at solving these identification problems.

Acknowledgements: We have benefitted from the comments of Melvin Adams, John Mullahy, and Gil Peled.

## 1. Introduction

Enormous attention and resources are being devoted to the quest for artificial general intelligence (AGI) and, even more ambitiously, artificial superintelligence. Much past research has described so-called *superhuman* artificial intelligence (AI) systems in specific contexts, such as a "superhuman AI program" for playing Go (Silver et al., 2017; Shin et al., 2023) and a "superhuman AI" poker player (Brown and Sandholm, 2019). The focus has recently shifted away from such narrow forms of AI to a much more general form that would outperform humans in a vast range of tasks and environments.

A 2024 article in *The Economist* titled "How to Define Artificial Intelligence" stated:[1] "Few definitions of AGI attract consensus…but most are based on the idea of a model that can outperform humans at most tasks—whether making coffee or making millions." Research on AGI is drawing attention well beyond academia and technology firms. Some commentators express optimism for solutions to pressing societal problems from cancer to climate change, while others warn of potentially severe consequences from mass unemployment to human extinction. Researchers already anthropomorphize current AI when they refer to errors committed by Large Language Models as "hallucinations."

Predictions of AGI and beyond are not new. Similar sentiments were expressed in 1960 by Herbert Simon, winner of the 1975 Turing Award for his pioneering research in artificial intelligence and human cognition[2] and the 1978 Nobel Prize in economics for research on decision-making within economic organizations.[3] Simon (1960) wrote: (p. 38):

> "Technologically, as I have argued, machines will be capable, within twenty years, of doing any work that a man can do. Economically, men will retain their greatest comparative advantage in jobs that require flexible manipulation of those parts of the environment that are relatively rough—some forms of manual work, control of some kinds of machinery (e.g., operating earth-moving equipment), some kinds of nonprogrammed problem-solving, and some kinds of service activities where face-to-face human interaction is of the essence."

---

[1] https://www.economist.com/the-economist-explains/2024/03/28/how-to-define-artificial-general-intelligence
[2] https://amturing.acm.org/award_winners/simon_1031467.cfm
[3] https://www.nobelprize.org/prizes/economic-sciences/1978/press-release/#:~:text=Professor%20Herbert%20A.,making%20process%20within%20economic%20organizations

We hear about these hopes and fears and wonder what the implications are for methodological research that aims to help decision makers cope with what econometricians call *identification problems*, inferential problems in empirical research that do not diminish as sample size grows. Of particular concern are missing data problems in prediction and treatment choice. Essentially all data collection intended to inform decision making is subject to missing data, which occurs for both mundane and fundamental reasons.

A leading mundane source of missing data in the social sciences is nonresponse to surveys. For example, reports of personal and household income have high rates of item nonresponse, in excess of 40 percent in the *Current Population Survey* (CPS); see Manski (2016). In addition to item nonresponse, researchers must cope with unit nonresponse. The U.S. Bureau of Labor Statistics reported the CPS household survey response rate to be 67.1 percent in July 2025, down from rates of close to 90 percent a decade earlier.[4] While these missing data rates are problematic, they compare very favorably with the norm in non-governmental surveys. For instance, recent election polls in the United States commonly have response rates below 2 percent (Dominitz and Manski, 2026a).

A fundamental source of missing data in policy analysis and medical research is the logical impossibility of observing counterfactual outcomes. This severely complicates analysis of treatment response and consequently complicates treatment choice. The problem is ubiquitous in research in the social sciences and medicine as well as in Silicon Valley, where so-called A/B testing has become prevalent. No one can observe the counterfactual outcome of treatment A for those who receive treatment B and, conversely, B for A. The unobservability of counterfactual outcomes is a problem often associated with analysis of observational data, but it inherently arises in randomized trials as well. Knowledge that treatments are randomized facilitates analysis of ideal trials, but ideal trials are rare in practice. Actual trials commonly have noncompliance and attrition, also known as loss to follow up.

[4] https://data.bls.gov/timeseries/LNU09300000&from_year=2013&output_type=column

Thus, coping with missing data problems is a central concern of decision making in public policy, health care, and many other fields. The methodological research literature has recommended and promoted many competing approaches to the problem, but there has been no silver bullet nor even any consensus regarding how best to proceed. It is essential to recognize that the problem cannot be solved by simply collecting more data in the same way. More survey respondents with the same rate of response does not solve the survey nonresponse problem. Larger trial size does not solve the problem of unobserved counterfactual treatment outcomes, nor those of noncompliance and attrition.

These problems can only be solved by bringing to bear information on the unobserved population of interest, identifying something about the incomes of those who do not report income or the treatment A outcomes for those who receive treatment B. Information that will reduce or ideally eliminate these identification problems can come in one of two forms: data or assumptions. On the former, rather than just more of the same data, it must be a different type of data, such as administrative records on the incomes of item non-respondents. On the latter, empirical researchers often make some sort of *missing at random* (MAR) assumption asserting that, for example, the income distribution among item non-respondents is identical to the distribution among item respondents who are similar in terms of other observed attributes.

A common approach to implementation of an MAR assumption is to impute data by drawing values at random from a specified distribution of observed data: the word "imputation" means using artificially constructed values, sometimes called "synthetic data," to take the place of missing data. See Rubin (1987). However imputation is performed, it necessarily uses assumptions about the distribution of missing data to generate the constructed values. The results depend critically on the assumptions made.

The central issue is the credibility of the maintained assumptions. The word *credibility* is in common use, but it has long defied deep definition. Aiming to provide a modicum of practical guidance to applied researchers regarding the assumptions that warrant their consideration, Manski (2003) counseled that they keep in mind a principle termed (p. 1):

> "*The Law of Decreasing Credibility*: The credibility of inference decreases with the strength of the assumptions maintained."

This principle implies that researchers face a dilemma as they decide what assumptions to maintain. Stronger assumptions yield stronger but less credible conclusions.

Researchers who purport to "solve" a missing data problem by using strong assumptions that lack credibility mislead themselves and their audiences. Manski (2011) calls this research with *incredible certitude*. The widespread adoption of MAR assumptions in empirical research represents a leading example of what Manski labels *conventional certitude*, whereby an assumption is generally adopted and accepted even if not correct, much like conventional wisdom. The seemingly less restrictive approach of Rubin (1977), which employs Bayesian thinking to formalize the problem that (p. 538) "the investigator has to make subjective judgments about the effect of nonrespondents" in surveys, also exhibits incredible certitude. It requires researchers to place a subjective prior distribution over a range of possible assumptions on missingness, even though there commonly is little credible basis to choose a prior.

Seeking to avoid incredible certitude, econometricians and other research methodologists have studied the identifying power of weak assumptions that partially identify the outcome distribution of interest. An example is a bounded-variation assumption that places an upper bound on the degree to which the presidential candidate preferences of poll non-respondents to election surveys differ from the preferences of respondents (Dominitz and Manski, 2026a). See Manski (2003) for a broad exposition.

Given the current enthusiasm about the potential of ongoing research in AI, we think it important to ask whether AI research is on a path to develop machines that outperform humans (that is, are superhuman) in addressing missing data problems, or perhaps even somehow solve these problems (that is, are superintelligent). Thus far, we see no indication that the current dominant architecture of machine learning (ML)-based AI systems will yield these advances. Moreover, we are concerned that the current mainstream ML approach to handling missing data may do more harm than good.

In this paper, we explain why we have reached these sobering conclusions and why we see the missing data problem as a cautionary case study in the quest for superintelligence more generally. Section 2 discusses the concept of intelligence, focusing first on some work by AI researchers, before presenting a decision-theoretic perspective that formalizes the connection between intelligence and identification problems. In Section 3, we apply this perspective to two leading cases of missing data problems. We explain in Section 4 why we are skeptical that we are currently on a path toward machines doing better than humans at solving these identification problems. Section 5 draws broader implications.

## 2. Concepts of Human and Artificial Intelligence

### 2.1. An Attempt to Define Universal Intelligence

It is natural to want to begin with an accepted definition of intelligence that enables comparison of human and artificial intelligence. But intelligence, as with so many superficially clear terms, has defied a consensus interpretation. Rather than take the space to review the vast multi-disciplinary literature on the subject, we think it instructive to consider a conscientious effort by two leading AI researchers to develop a definition of *universal intelligence*, which would be applicable to humans, animals, and machines.

At the outset, the authors Shane Legg and Marcus Hutter, who have recently been senior researchers at Google DeepMind, observe (Legg and Hutter, 2007, p. 391): "A fundamental problem in artificial intelligence is that nobody really knows what intelligence is." In the first half of their lengthy article, they give an extensive review of over a century of psychological research on human intelligence, citing many distinct verbal definitions in the literature. They eventually propose their own verbal definition, writing (p. 402):

> "Bringing these key features together gives us what we believe to *[be]* the essence of intelligence in its most general form: Intelligence measures an agent's ability to achieve goals in a wide range of environments."

This definition is similar to many that relate intelligence broadly to performance in decision making. For example, in the mid-1900s the psychologist David Weschler defined intelligence as follows (Wechsler, 1958, p. 7): "The aggregate or global capacity of the individual to act purposefully, to think rationally and to deal effectively with his environment."

In the second half of the article, Legg and Hutter develop a mathematical definition of universal intelligence as a scalar measure of performance in decision making. They aim to formalize the word "universal" in two senses. One is to measure decision performance in "a wide range of environments." The other is that the agent making decisions is an abstract entity, which could be either a human or a machine.

To guide them in this challenging task, the authors place considerable stock in the medieval principle of Occam's razor, which they cite as (p. 412): "Given multiple hypotheses that are consistent with the data, the simplest should be preferred." They state: "This is generally considered the rational and intelligent thing to do." More recently, in their book *An Introduction to Universal Artificial Intelligence*, Hutter et al. (2024) described Occam's razor as (p. 8): "The most important philosophical tool for science."

With Occam's razor in mind, Legg and Hutter bring to bear a particular type of Bayesian thinking. Considering allegedly simple real-world environments to be more likely in a specific way than allegedly complex environments, they place what they call an *algorithmic probability distribution* across all environments that a decision maker might possibly face. They then propose measurement of universal intelligence as the likelihood-weighted expected performance of a decision maker across these environments, with more complex environments assigned lower algorithmic probabilities than simpler ones.

From the perspective of our interest in decision making with missing data, we welcome the effort of Legg and Hunter to relate intelligence to performance in decision making in a wide range of environments. Relatedly, Manski (2004) and Dominitz and Manski (2017, 2026b), among others, have advocated use of statistical decision theory (Wald, 1939, 1945, 1950) to coherently evaluate decision making in a wide range of environments. However, we cannot sympathize with their appeal to Occam's razor to motivate their mathematical definition of universal intelligence. Manski (2011, 2020) has cautioned against attempts to apply this principle. We paraphrase and expand on the argument below.

2.1.1. Misguided Appeals to Occam's Razor

Scientists often refer to *Occam's Razor*, the medieval philosophical declaration that "Plurality should not be posited without necessity." Duignan (2025), writing in the *Encyclopaedia Britannica*, gives the usual modern interpretation of this cryptic statement, remarking that: "The principle gives precedence to simplicity; of two competing theories, the simplest explanation of an entity is to be preferred." The choice criteria offered by Legg and Hunter and by Duignan are as imprecise as the one given by Occam. What is meant by "simplicity?"

In an influential methodological essay, Milton Friedman placed prediction as the central objective of science, writing (Friedman, 1953, p. 5): "The ultimate goal of a positive science is the development of a 'theory' or 'hypothesis' that yields valid and meaningful (i.e. not truistic) predictions about phenomena not yet observed". He went on to say (p. 10):

> "The choice among alternative hypotheses equally consistent with the available evidence must to some extent be arbitrary, though there is general agreement that relevant considerations are suggested by the criteria 'simplicity' and 'fruitfulness,' themselves notions that defy completely objective specification."

Thus, Friedman counseled scientists to use Occam's razor choose one hypothesis, even though this may require the use of "to some extent... arbitrary" criteria. He did not explain why scientists should choose a single hypothesis out of many. He did not entertain the idea that scientists might offer predictions under the range of plausible hypotheses that are consistent with the available evidence.

Occam's razor appears to have long played a role in statistical learning theory and machine learning research, as described recently by Sterkenburg (2025). Reliance on the concept dates at least as far back as Pearl (1978), who asserted (p. 255) that the "human tendency to regard the simpler as the more trustworthy is given a qualified justification," and Blumer et al. (1987), who defined an "Occam-Algorithm" and described (p. 377) an "emerging science of machine learning, whose expressed goal is often to discover the simplest hypothesis that is consistent with the sample data."

These ML arguments for simplicity arose from pragmatic concerns related to computational complexity and the generalizability of predictions outside of the training data, rather than from the assertion that a simpler model or environment is more likely to be true than a more complex one. Sterkenburg (2025) concluded his "core 'means-ends' argument" for simplicity as follows (p. 24):

> "In this paper, I described a means-ends justificatory argument for Occam's razor from statistical learning theory…It is an honest epistemic justification; and *it does not rely on any extra-theoretical assumptions about the true or best hypothesis being simple*" *(italics added for emphasis)*.

We agree that simplicity may be a well-defined and relevant concept when considering algorithms that vary in their computational complexity. However, the broader relevance of Occam's razor to decision making is obscure. The word "simplicity" is rather vague. It appears that humans differ enormously in how they interpret the word. We conjecture that hypothetical superintelligent machines might differ as well in their interpretations. However one may define simplicity, we are not aware of a serious foundation for the belief that simpler environments or hypotheses are more likely to be true than more complex ones.

Indeed, science gives reason to think that simplicity and complexity co-exist throughout our physical, biological, and social universe. Elementary particles combine to form elements, molecules, planets, and galaxies. Binary chips connect to form computers and the internet. Organic molecules combine to form cells, plants, and mammals. Individual humans combine to form families, communities, firms, and nations.

## 2.2. A Decision Theoretic Perspective on Intelligence

We find it productive to consider intelligence within the mathematical framework of decision theory. As far as we are aware, psychological research on intelligence has not used this framing. Computer science research on AI has sometimes used decision-theoretic concepts, but it has not systematically brought the theory to bear as we do here.

Decision theory has long been used widely in economics, statistics and operations research, as well as recently in ML research, to structure normative and prescriptive analysis of decision making. Some central ideas have roots as far back as the 1700s, with general formalization developing in the 1900s. Wald's statistical decision theory, mentioned above, extends earlier decision theory to encompass learning from sample data. The primary problems in decision making with missing data arise in mathematically less challenging settings where information is deterministic rather than generated by sampling. For expositional clarity, we will not discuss the Wald theory here.

In what follows, we first briefly summarize the most basic features of decision theory. Among the numerous textbook expositions, we suggest Berger (1985) and Ferguson (1967) for those who want to learn more. We then discuss how intelligence may matter.

2.2.1. Decision Theory

Decision theory posits a decision maker (DM) who faces a predetermined choice set. The decision maker ideally wants to choose a feasible action that maximizes a specified welfare function (equivalently, minimizes a loss function). However, the DM has incomplete knowledge of the environment that is faced. Hence, the DM faces a problem of choice under uncertainty.

Formally, let C denote the choice set. Let S denote a specified set of possible environments, usually called *states of nature*, and let $s^*$ denote the unknown true state. The state space S may be finite-dimensional (parametric) or larger (nonparametric), but it must be specified ex ante by the DM. In colloquial terms, decision theory studies choice among "known unknowns" rather than "unknown unknowns," with S expressing the DM's knowledge of the possible states.

A specified objective function $w(\cdot, \cdot)$ maps actions and states into a real-valued welfare. The DM ideally would maximize $w(\cdot, s^*)$ over C, but this is not achievable because the DM does not know $s^*$. To cope with uncertainty, the DM proceeds in two steps.

The first step is to eliminate dominated actions from consideration. A feasible action c is said to be weakly dominated if there exists another one d such that $w(d, s) \geq w(c, s)$ for all possible states of nature s and $w(d, s) > w(c, s)$ for some s. There is essentially consensus among researchers that dominated actions should be eliminated, provided that they can be determined without cost. To intentionally choose a dominated action would be, well, unintelligent.

The second step is to choose among the actions that are undominated, or at least not known to be dominated. This step is fundamentally difficult, at least from the perspective of human intelligence. There is no singularly optimal way to proceed, there at most are "reasonable" ways. Ferguson (1967) put it this way (p. 28):

> "It is a natural reaction to search for a 'best' decision rule, a rule that has the smallest risk no matter what the true state of nature. Unfortunately, *situations in which a best decision rule exists are rare and uninteresting*. For each fixed state of nature there may be a best action for the statistician to take. However, this best action will differ, in general, for different states of nature, so that no one action can be presumed best overall."

To choose an action in an arguably reasonable way, decision theorists have proposed using the welfare function to form functions of actions alone, which can be optimized. Perhaps most widely discussed is *Bayesian decision theory*, which places a subjective probability distribution π on the state space, computes average state-dependent welfare with respect to π, and maximizes subjective expected welfare over C. The criterion solves the maximization problem

$$(1) \qquad \max_{c \in C} \int w(c, s) d\pi.$$

The universal intelligence function developed by Legg and Hutter has this form, with π being their algorithmic probability function.

Other approaches avoid specification of a subjective probability distribution and instead seek an action that, in some sense, works uniformly well over all of S. The most prominent expressions of this idea are the maximin and minimax-regret criteria. Wald (1950) studied the *maximin* criterion, which considers the worst that can be happen with choice of each action. It solves the problem

$$(2) \qquad \max_{c \in C} \; \min_{s \in S} \; w(c, s).$$

Savage (1951), in a book review of Wald (1950), criticized the pessimism of considering only worst-case states of nature and suggested a different formalization of the idea of selecting an action that works uniformly well over all of S. This formalization, which has become known as the *minimax-regret* (MMR) criterion, solves the problem

$$(3) \quad \min_{c \in C} \max_{s \in S} [\max_{d \in C} w(d, s) - w(c, s)].$$

Here $\max_{d \in C} w(d, s) - w(c, s)$ is called the *regret* of action c in state s. The true state being unknown, one evaluates c by its maximum regret over all states and selects an action that minimizes maximum regret.

The maximum regret of an action measures its maximum distance from optimality across states. Manski (2004, 2021) and Dominitz and Manski (2017, 2026b) argued that this is an appealing feature of using the MMR criterion when applying statistical decision theory. Furthermore, in the familiar context of prediction under a square loss function, maximum regret corresponds to the familiar measure of maximum mean square error (MSE).

2.2.2. Intelligent Specification of the State Space

Decision theory considers any DM, human or machine, who uses the above formal structure to make decisions to behave reasonably. The theory does not assert that application of any of the decision criteria we have described—Bayes, maximin, minimax-regret—to be more intelligent than the others. Nor does the theory take a stand on what welfare function a DM should want to optimize. This is viewed as a meta-choice made before contemplating choice of an action, expressing the personal preferences of the DM. Some observers may consider certain preferences to lack common sense or to be unethical, but decision theory does not question preferences.

Where then does intelligence come into play? One might construe a narrow sense of intelligence to be a computational capacity to determine dominated actions and to solve the mathematical problems (1) through (3). We do not adopt this perspective. To understand why, we quote again from Herbert Simon, who had the computational limits of humans in mind in the article that spawned the modern literature in behavioral economics, writing (Simon, 1955, p. 101):

> "Because of the psychological limits of the organism (particularly with respect to computational and predictive ability), actual human rationality-striving can at best be an extremely crude and simplified approximation to the kind of global rationality that is implied, for example, by game-theoretical models."

Simon and the literature that followed him have not described the computational limits of humans as a lack of intelligence. The term *bounded rationality* has been used. We agree that the term intelligence does not fit. After all, humans have long augmented their innate computational capacities by inventing machines to assist them. An abacus, slide rule, hand calculator, or computer that performs computations faster and more accurately than a human is not more intelligent than humans. It is just a device used by humans. When AI researchers write of artificial superintelligence, they mean much more than computational prowess. They have in mind crossing a so-called "singularity" in AI development after which AI will be capable of unending self-improvement with no need for human intervention.

One might construe an aspect of intelligence to be awareness of the full range of options available in the choice set. Some research in marketing and behavioral economics has hypothesized that humans may choose actions from "consideration sets," which are subjectively determined subsets of choice sets. Research using the concept of a consideration set has not achieved a consensus about the cognitive process that may yield this phenomenon. Some investigators conjecture that boundedly rational DMs intentionally restrict attention to consideration sets, recognizing that they do not have the computational capacity to evaluate all feasible options (e.g., Hauser, 2014). If so, we would not necessarily interpret choice from consideration sets to indicate a lack of intelligence. Rather, we would connect use of consideration sets to intelligence if a DM can costlessly recognize and evaluate a complete choice set, yet disregards some options.

We view the general reasoning ability conveyed by intelligence to appear in decision theory in the specification of the state space. As an abstraction, the state space of decision theory is a subjective precursor to decision making, a primitive concept that expresses uncertainty. The larger the state space, the less the DM knows about the consequences of each action.

In principle, decision theory can be applied with any specification of the state space. However, the specification matters. Holding fixed the (choice set, welfare function, decision criterion) triple, the chosen action can vary markedly with the specified state space.

In practice, humans clearly do not consider all state spaces to be created equal. When facing a particular decision setting, humans often express heterogeneous views regarding the extent and nature of available knowledge. At one extreme, an individual may have a strong but unsubstantiated belief regarding the true state of nature, taking the state space to contain a single element. Manski (2011) used the term *incredible certitude* to describe a strong individual belief that lacks a credible foundation. The term *dueling certitudes* describes a situation in which different individuals express competing incredible certitudes. At another extreme, an individual may perceive unrealistically huge uncertainty about the true state of nature, taking the state space to be a very large set. We might use the word *nihilist* to describe someone who makes decisions that disregard credible information.

Broadly speaking, we might regard intelligence to be a general ability to specify a realistic state space. The state space should manifest neither incredible certitude, which assumes erroneous information about the environment, nor nihilism, which fails to use credible information. Decision making with a smaller than realistic state space erroneously classifies some actions as dominated. Use of a larger than realistic state space erroneously classifies some actions as undominated. Neither appropriately measures subjective expected welfare, minimum welfare, or maximum regret.

The central open issue in the above paragraphs is interpretation of the words "credible" and "realistic." Simply stating that an intelligent DM should specify a credible or realistic state space is meaningless per se. It just transfers vagueness in the definition of intelligence to vagueness about the meaning of credibility and realism.

In Bayesian decision theory, where the DM places a subjective probability distribution on the state space, a subset of the state space that has high subjective probability is called a *credible set*. In this terminology, states of nature outside of the state space have zero credibility. The Bayesian idea of a credible set is well-defined once one has specified the state space, but it is unattractive when a DM seeks to compare different choices for the state space. Suppose that one initially specifies a small state space and then considers enlarging it to consider possibilities not previously recognized. Subjective probabilities must sum to one. Hence, expansion of the state space necessarily reduces the Bayesian credibility of the states in the initial state space.

Research in abstract decision theory has been silent on specification of the state space, considering it to be entirely subjective. However, research in the sciences seeks to provide at least a partially objective basis for specification in particular contexts, obtained by combining well-motivated assumptions (aka theory) with empirical analysis of available data to generate reliable information about the real world. Research methodologists, including econometricians such as us, study the conclusions that logically may be drawn by combining various types of assumptions and data. When these conclusions are considered to be deterministic, they yield the state space. (Wald's statistical decision theory addresses the more subtle problem of decision making with sample data, where conclusions informed by scientific research are not deterministic.)

*Identification Analysis and the State Space*

It has been standard in econometrics to specify the state space as a set of objective probability distributions that may possibly describe the system under study. Haavelmo (1944) did so for economic systems when he introduced *The Probability Approach in Econometrics*. Studies of treatment choice do so when they consider the population to be treated to have a distribution of treatment response.

The Koopmans (1949) formalization of identification analysis contemplated unlimited data collection that enables one to shrink an initially specified state space, eliminating states that are inconsistent with accepted theory and with the information revealed by observation of data. For most of the 20$^{th}$ century, econometricians commonly thought of identification as a binary event – a feature of an objective probability distribution (a parameter) is either identified or it is not. Empirical researchers applying econometric methods combined available data with assumptions that yield point identification, in which case the state space contains only one element. Economists recognized that point identification often requires strong assumptions that are difficult to motivate. However, they saw no other way to perform empirical research.

Yet there is enormous scope for fruitful research using weaker and more credible assumptions that partially identify population parameters. A parameter is partially identified if the sampling process and maintained assumptions reveal that the parameter lies in a set, its *identification region* or *identified set*, that is smaller than the logical range of the parameter but larger than a single point. In econometrics, the terms identification region and identified set are synonyms for the state space. Thus, econometric analysis of identification aims to determine a realistic state space.

Isolated contributions to analysis of partial identification were made as early as the 1930s, but the subject remained at the fringes of econometric consciousness and did not spawn systematic study. A coherent body of research took shape in the 1990s and has since grown rapidly. Reviews of this work include Manski (1995, 2003, 2007), Tamer (2010), and Molinari (2020).

Recognizing the Law of Decreasing Credibility, econometricians studying partial identification have recommended that applied researchers perform a sensitivity analysis that systematically explores the tradeoff between the identifying power and the credibility of assumptions, seeking to learn a balance that they find acceptable. One might first determine the conclusions that can be drawn with minimal assumptions and then progressively strengthen them. Or one might begin with strong assumptions that yield point identification and then weaken them. Either way, the cognitive process of exploring a range of assumptions can be enlightening, we dare say intelligent.

Partial identification analysis was first connected to decision theory in Manski (2000), writing (p. 416):

> "This paper connects decisions under ambiguity with identification problems in econometrics. Considered abstractly, it is natural to make this connection. Ambiguity occurs when lack of knowledge of an objective probability distribution prevents a decision maker from solving an optimization problem. Empirical research seeks to draw conclusions about objective probability distributions by combining assumptions with observations. An identification problem occurs when a specified set of assumptions combined with unlimited observations drawn by a specified sampling process does not reveal a distribution of interest. Thus, identification problems generate ambiguity in decision making."

The terminology in the above paragraph follows Ellsberg (1961) in using the word *ambiguity* to signify uncertainty when one specifies a set of feasible states of nature but does not place a probability distribution on the state space as in Bayesian analysis. Synonyms for ambiguity include *deep uncertainty* and *Knightian uncertainty*.

## 3. Prevalent Approaches to Coping With Missing Data

### 3.1. General Considerations

The discussion of identification at the end of Section 2 leads naturally to our concern for decision making with missing data. Identification is the primary difficulty created by missing data.

To understand the problem, consider the common scenario in public policy or clinical decision making in which a DM must choose treatments for the members of a population. The optimal treatment rule depends on the population distribution of individual covariates and treatment outcomes. To learn about this distribution, members are sampled at random, but the values of relevant outcomes and/or covariates are not observed for some sampled members.

Drawing a larger sample will not help—missing data in the initial sample remain missing, and some members of the larger sample will inevitably also have missing data. Thus, missing data is not primarily a problem of statistical imprecision that disappears as sample size goes to infinity. The problem primarily is that one can learn the distribution of values only for the sub-population who provide data, not for the complementary sub-population whose values are unobserved. This is an identification problem.

Wanting to achieve point identification, shrinking the state space to one distribution of missing data, empirical researchers in the social sciences, medicine, and other fields have commonly maintained some version of the assumption that data are missing at random (MAR), in the sense that the observability of a variable is statistically independent of its value. Yet this and other point-identifying assumptions have regularly been criticized as implausible. Thus, research assuming that data are MAR commonly suffers from incredible certitude.

In contrast, study of partial identification begins with an agnostic analysis that determines what the data generation process reveals about the relevant population if nothing is known about the distribution of missing data. One then brings to bear credible weak assumptions and determines their identifying power. It is commonly the case that such assumptions shrink the state space but do not reduce it to a point. Thus, the quest for a realistic state space commonly yields a set of possible distributions for the missing data, not a single distribution. The state space (aka identification region), which depends on the maintained assumptions, is this set.

A severe difficulty in human research has been absence of agreement on what assumptions to maintain for the distribution of missing data. Researchers who seek point identification vary in what economists call their *identification strategies*; that is, the assumptions they use to shrink the state space to a point. Researchers who study partial identification also vary in the assumptions they make. Agnostic analysis that places no restrictions on the distribution of missing data is sometimes considered nihilistic; that is, more conservative than is realistic. Yet researchers vary in the assumptions that they deem sufficiently credible to warrant using them. Agnostic analysis is not nihilistic when researchers have little understanding of the process yielding missing data.

To explain more concretely, we discuss approaches to two common research problems. Section 3.2 addresses conditional prediction with missing outcome data. Section 3.3 describes how researchers seek to cope with missing data on counterfactual outcomes in analysis of treatment response.

We describe these problems in some detail for two reasons. One is to clarify key considerations for intelligent application of credible, context-dependent assumptions that may shrink the state space. The other is to provide a rigorous foundation for our assertion in Section 4 that, with the current ML-based AI architecture, machines will not do better than humans at developing credible assumptions to solve the identification problem created by missing data.

3.2. Conditional Prediction with Missing Outcome Data

A longstanding concern of statistics and econometrics has been development of methods to use observable data to predict an outcome y conditional on specified covariates x. For example, a biostatistician or health economist performing research that aims to inform medical decision making may want to predict whether a person with health history and demographic attributes x will develop a specified illness (y =1 if yes, y = 0 if no) or, perhaps, will live for y years.

A standard formalization considers a heterogeneous population characterized by a joint distribution P(y, x), where y is a real outcome and x is a covariate vector. The objective is to learn about the conditional distribution P(y|x). Ideally, one observes ($y_i$, $x_i$, i = 1, . . , N) in a random sample of N persons drawn from a study population that has distribution P(y, x). One uses the sample data to estimate features of P(y|x). Research has particularly focused on the conditional mean E(y|x) or median M(y|x). These are the best predictions of y under square and absolute loss, respectively, properties which provide decision-theoretic motivations for them.

Incomplete observability of sample data generates an identification problem. Agnostic inference contemplates all logically possible distributions of the missing data. Doing so yields the set of all possible values of P(y|x), its identification region. Assumptions about the distribution of missing data have identifying power. Weak assumptions may shrink the identification region for P(y|x). Sufficiently strong assumptions may yield point identification.

A practical challenge is to characterize the identification region in a tractable way. Manski (1989, 1994) showed that identification analysis for E(y|x) and conditional quantiles is elementary when only outcome data are missing. Analysis is more complex when the objective is to learn a spread parameter such as Var(y|x); see Blundell et al. (2007) and Stoye (2010). Analysis is also more complex when sample members have missing covariate data. Horowitz and Manski (1998, 2000), Manski (2018), and Venkataramani, Manski, and Mullahy (2026) study these settings, with focus on E(y|x).

In this space, we formalize the identification problem in an important setting without mathematical complexity. Consider identification of the conditional mean $E(y|x)$ when only outcome data are missing and y is a bounded outcome, whose measurement is normalized so that y takes values in the interval [0, 1]. An elementary argument presented in Manski (1989) yields the identification region for $E(y|x = \xi)$ for any value of x, say $\xi$, that occurs with positive probability in the population.

For each member of the population, let $z = 1$ indicate whether y is observable and $z = 0$ otherwise. Thus, missing data on y occur when $z = 0$. The Law of Iterated Expectations gives

$$(4) \quad E(y|x = \xi) = E(y|x = \xi, z = 1)P(z = 1|x = \xi) + E(y|x = \xi, z = 0)P(z = 0|x = \xi).$$

Among the quantities on the right-hand side, $E(y|x = \xi, z = 1)$ and $P(z = 1|x = \xi)$ are point-identified and can be estimated consistently by observing a random sample of the population. However, nothing is empirically learnable about $E(y|x = \xi, z = 0)$, the mean outcome in the sub-population with missing data. Agnostic identification analysis recognizes only that $E(y|x = \xi, z = 0)$ must lie in the interval [0, 1]. Hence, the agnostic identification region for $E(y|x = \xi)$ is the interval

$$(5) \quad [E(y|x = \xi, z = 1)P(z = 1|x = \xi),\ E(y|x = \xi, z = 1)P(z = 1|x = \xi) + P(z = 0|x = \xi)].$$

This interval has width $P(z = 0|x = \xi)$, the fraction of the population whose outcomes are not observable.

Suppose that the researcher maintains assumptions that restrict $E(y|x = \xi, z = 0)$ to a proper subset of [0, 1], say $\Gamma$. Returning to the Law of Iterated Expectations, the identification region is

$$(6) \quad E(y|x = \xi, z = 1)P(z = 1|x = \xi) + \gamma \cdot P(z = 0|x = \xi),\ \gamma \in \Gamma.$$

$E(y|x = \xi)$ is point-identified if the maintained assumptions imply that $E(y|x = \xi, z = 0)$ must take a specific value. Suppose, for example, that a researcher assumes the data are MAR. Then $E(y|x = \xi) = E(y|x = \xi, z = 1)$).

In what follows, Section 3.2.1 critiques the widespread use of MAR assumptions in empirical research on conditional prediction. Section 3.2.2 critiques selection modeling, which has been the main point-identifying alternative to the MAR assumption. Section 3.2.3 calls attention to bounded-variation assumptions. These weaken MAR assumptions, increasing credibility at the expense of identifying power. Section 3.2.4 uses election polling to illustrate.

### 3.2.1. MAR Assumptions

The *mean-independence* form of the MAR assumption, namely $E(y|x = \xi) = E(y|x = \xi, z = 1)$, and its stronger *statistical independence* form, $P(y|x = \xi) = P(y|x = \xi, z = 1)$, have long been used in empirical research on conditional prediction. Early on, researchers often coped with missing data by directly assuming that $E(y|x = \xi) = E(y|x = \xi, z = 1)$, without reference to the sub-population $P(y|x = \xi, z = 0)$ with missing data. Missingness was said to be *ignorable*. The MAR assumption is very simple, so Occam's razor gives it a surface appeal.

In the 1970s, statisticians and econometricians independently raised awareness that an MAR assumption may not be credible. After all, given any specification of (y, x), the MAR assumption picks out one particular distribution of missing data from all that are logically possible, regardless of context. The statistician Donald Rubin popularized the term *missing at random* in Rubin (1976) and in numerous subsequent publications. In econometric research, the assumption was commonly called *selection on observables*; Fitzgerald, Gottschalk, and Moffitt (1998, Section IIIA) discuss the history. The MAR terminology eventually became prevalent.

Statisticians and econometricians agreed that the assumption has a highly credible foundation in settings where missingness is known to arise from a well-understood random process. The most famous is missingness of data on counterfactual outcomes in ideal randomized controlled trials (RCTs), where treatments are assigned randomly; see Section 3.3 for further discussion. The two disciplines developed sharply contrasting perspectives on the credibility of MAR assumptions in observational settings such as occur in survey research or in analysis of treatment response when treatments are not assigned randomly.

In observational settings, econometricians were largely skeptical of the credibility of MAR assumptions, arguing that missingness of outcome data is often determined by personal choices that may vary with personal outcomes. A leading example was in prediction of wages in the labor market. Widely accepted economic theory posited that individuals choose to work if their market wage is above a person-specific threshold called a *reservation wage*, and they choose not to work otherwise. Hence, market wages are observable only when they exceed the reservation wage. Gronau (1974) called this phenomenon a *selectivity bias* (aka *selection bias*). Economists similarly conjectured that self-selection of treatments in settings outside of ideal RCTs would often falsify MAR assumptions in analysis of treatment response. The reasoning was that individuals would choose treatments that yield more favorable outcomes, so observed outcomes would tend to be more favorable than counterfactual outcomes.

Notwithstanding the choice-related arguments of economists, Rubin and other statisticians have argued that the MAR assumption becomes increasingly credible as one conditions prediction on increasingly many covariates. Rubin (1977) offered the following justification in the context of survey nonresponse (p. 540):

> "As more background variables are recorded, the investigator should be more confident that if the *Y* means for respondents and nonrespondents differ, the difference should be reflected in the observed distribution of the background variables…because as more background variables are recorded, it becomes more likely (to be precise, not less likely) that the propensity to be a nonrespondent is some probabilistic function of the recorded background variables."

Roderick Little, a frequent collaborator of Rubin, has written (Little, 2021, p. 102): "Rubin himself has argued that with a sufficiently rich set of observed data, MAR is often justified." Little also wrote: "The problem is that we usually cannot tell from the observed data whether or not MAR applies."

Regarding this belief in the power of additional covariates, Manski (2007) argued (pp. 65-66):

> "Researchers often assert that missingness at random conditional on (x, w) is more credible than is missingness at random conditioning on x alone. To justify this, they say that (x, w) 'controls for' more

> determinants of missing data than does x alone. Unfortunately, the term 'controls for' is a vague expression with no formal standing in probability theory. When researchers say that conditioning on certain covariates 'controls for' the determinants of missing data, they rarely give even a verbal explanation of what they have in mind, never mind a mathematical explanation.
>
> There is no general foundation to the assertion that missingness at random becomes a better assumption as one conditions on more covariates. The assertion may be well grounded in some settings, but it is not self-evident. Indeed, outcomes may be missing at random conditional on x but not conditional on the longer covariate vector (x, w)."

Manski (2007, Section 2.6) gave an example based on the reservation-wage model of labor supply.

### 3.2.2. Selection Modeling

Being skeptical of MAR assumptions, but wanting to achieve point-identification, econometricians of the 1970s developed parametric *selection models* that aim to explain non-randomly missing data by choice processes that render some outcomes observable and others not. Prominent contributions include Gronau (1974) and Heckman (1979). Maddala (1983) provides an extensive exposition, with explanation of assumptions that are necessary and sufficient for point-identification and consistent estimation of the parameters. After an initial period of enthusiasm, the credibility of these assumptions was increasingly questioned. From the 1980s onward, econometricians have sought to weaken the assumptions while retaining point-identification, developing various semiparametric and nonparametric models. See Blundell and Powell (2003).

Doubts about the credibility of selection models are justifiable. This does not imply, however, that MAR should be the preferred alternative. To the contrary, we think it instructive to quote the concluding paragraph from the seminal MAR paper by Rubin (1976), which is contemporaneous with the early econometric work on modelling missingness (p. 589):

> "The inescapable conclusion seems to be that when dealing with real data, the practicing statistician should explicitly consider the process that causes missing data far more often than he does. However, to do so, he needs models for this process and these have not received much attention in the statistical literature."

3.2.3. Bounded-Variation Assumptions

With agnostic analysis of missing data at one pole and point-identifying assumptions such as MAR at the other, a researcher can contemplate a vast spectrum of assumptions that shrink the agnostic identification region for $E(y|x = \xi)$ but are not strong enough to yield point identification. Particularly intuitive are *bounded-variation* assumptions, which constrain the distance between the observable conditional mean $E(y|x = \xi, z = 1)$ and the unobservable one $E(y|x = \xi, z = 0)$, Formally, these assumptions have the form

(7) $$\delta_0 \leq E(y|x = \xi, z = 1) - E(y|x = \xi, z = 0) \leq \delta_1,$$

where $\delta_0$ and $\delta_1$ are specified positive constants. Bounded-variation assumptions have been applied in Manski (2018), Manski and Pepper (2018), Li, Litvin, and Manski (2023), Dominitz and Manski (2026a), and elsewhere.

Identification analysis with bounded-variation assumptions is straightforward. The tighter the constraint on the distance between the observable and unobservable mean, the greater is the identifying power. Assessment of the credibility of an assumption must be context-specific. The applications cited above bring to bear available information about the context to motivate the assumptions imposed.

3.2.4. Illustration: Election Polling

The well-known context of election polling is illustrative. Much attention in the months preceding a national election is devoted to poll results that form the basis for point predictions of the election outcome. As noted in the Introduction, recent election polls in the United States commonly have response rates below 2 percent. Pollsters often acknowledge concerns about these response rates. Prosser and Mellon (2018) observed that polling analysts typically weight the available data to attempt to correct for non-response bias, noting (p. 772): "Survey researchers have long known that the ability of weighting to correct for survey bias rests on the assumption that respondents mirror non-respondents within weighting categories." That is, weighting approaches assume that responses are MAR. Bailey (2023) called for selection modeling rather than weighting, based on the belief that the choice to respond to a poll may vary with candidate preferences, conditional on observed attributes x.

Conventional reporting of poll results aim to minimize mean square error. Assuming that nonresponse is random, pollsters ignore bias and focus on variance. Considering this setting—prediction under square loss in the presence of survey non-response—Dominitz and Manski (2017) studied minimax regret prediction when response may not be MAR. Dominitz and Manski (2026a) applied the methodological findings to election polling. We draw on that work here.

Dominitz and Manski (2026a) began with an agnostic analysis to determine what the data reveal about the relevant population if nothing is known about the candidate preferences of non-respondents. With non-response rates in excess of 98 percent, the short answer is "not much." For example, the estimated identification region (5) for the preference for the Republican candidate (Donald Trump) in May 2024 was [0.007, 0.994], given the expressed support of 54.4% of poll respondents and a response rate below 1.4%.

In contrast, assuming data are MAR, the preference for Trump is point-identified and estimated to be 0.544. But the MAR assumptions conventionally made by pollsters are not credible. Their assertions of point identification manifest incredible certitude.

The analysis instead considered bounded-variation assumptions of the form (7). The central question then becomes how to determine credible values for $\delta_0$ and $\delta_1$, the bounds on the difference in preferences between respondents $P(y = 1|x = \xi, z = 1)$ and non-respondents $P(y = 1|x = \xi, z = 0)$. It may seem attractive to use historical evidence to determine credible values. However, the credibility of using historical evidence depends on the quality of the evidence and on how one assesses the stability over time of the election environment. To illustrate the issues, Dominitz and Manski discussed a study by Shirani-Mehr et al. (2018), who examined 4221 state-level polls conducted in the final three weeks before presidential, senatorial, and gubernatorial elections from 1998 through 2014.

To intelligently shrink the state space using bounded-variation assumptions requires determining credible, context-dependent values for $\delta_0$ and $\delta_1$. This might be possible if pollsters are able to combine follow-up studies of non-respondents with analysis of election results, seeking to learn how the preferences of non-respondents tend to differ from respondents.

### 3.3. Missing Data on Counterfactual Outcomes in Prediction of Treatment Response

Prediction of treatment response poses a pervasive and distinctive problem of prediction with missing outcomes. Studies of treatment response aim to predict the outcomes that would occur if alternative treatment rules were applied to a population. One cannot observe the outcomes that a person would experience under all treatments. At most, one can observe a person's *realized outcome*; that is, the one experienced under the treatment actually received. The *counterfactual outcomes* that a person would have experienced under other treatments are logically unobservable. Thus, missing data is inevitable in prediction of treatment response.

A standard formalization of the prediction problem considers a population whose members have observed covariates denoted x. A set of mutually exclusive and exhaustive alternative treatments is denoted T. For each $t \in T$, y(t) is a real-valued outcome that the person would experience with treatment t. It is usually assumed that treatment is individualistic; that is, the treatment received by one person does not affect the outcomes experienced by others. The general objective is to learn the conditional distributions of treatment outcomes $P[y(t)|x]$, $t \in T$. A specific objective often is to learn the conditional mean outcomes $E[y(t)|x]$, $t \in T$.

Let z denote the treatment received by a member of the population. Let $P(z = t|x = \xi)$ be the fraction of persons in the population who receive t, among those with covariate value $x = \xi$. Let y denote a person's observable realized outcome, which is y(t) when $z = t$. The Law of Iterated Expectations and the fact that $y(t) = y$ when $z = t$ give

$$(8) \quad E[y(t)|x = \xi] \; = \; E(y|x = \xi, z = t)\cdot P(z = t|x = \xi) + E[y(t)|x = \xi, z \neq t]\cdot P(z \neq t|x = \xi).$$

Here $E[y(t)|x = \xi, z = t] = E(y|x = \xi, z = t)$ is mean treatment response within the group who have covariates $\xi$ and who receive treatment t, whereas $E[y(t)|x = \xi, z \neq t]$ is mean response for those who receive another treatment. Abstracting from statistical imprecision, observation of realized treatments and outcomes in a random sample of the population reveals $P(z = t|x = \xi)$ and $E[y(t)|x = \xi, z = t]$. The distribution $E[y(t)|x = \xi, z \neq t]$ is counterfactual, hence unlearnable from observation.

Replacing y and $z \in \{0, 1\}$ of Section 3.2 with y(t) and $z \in T$, equations (4) and (8) are equivalent. If y(t) has the bounded range [0, 1], the agnostic identification region for $E[y(t)|x = \xi]$ is analogous to (5), namely

$$(9) \quad [E(y|x = \xi, z = t)P(z = t|x = \xi),\ E(y|x = \xi, z = t)P(z = 1|x = \xi) + P(z \neq t|x = \xi)].$$

The width of this interval is the fraction of persons in the population who do not receive treatment t; that is, the fraction for whom y(t) is counterfactual.

As in Section 3.2, the MAR assumption $E[y(t)|x, z = t] = E[y(t)|x, z \neq t]$ point-identifies $E[y(t)|x]$. This assumption has unquestioned credibility in ideal randomized experiments, where it is known that treatments are received randomly. However, ideal randomized experiments are rare. The MAR assumption may not hold in realistic experiments where some subjects do not comply with assigned treatments. The credibility of the assumption may be minimal in *observational studies*, where realized treatments are consciously chosen by members of the population, whose choices may be related to their treatment outcomes.

Bounded-variation assumptions weaken the MAR assumption by bounding the difference between $E[y(t)|x, z = t]$ and $E[y(t)|x, z \neq t]$. As discussed in Section 3.2.3, these assumptions can increase credibility, but they achieve only partial identification of mean treatment response. Thus, the Law of Decreasing Credibility must be respected.

Concerned with noncompliance in experiments and with conscious choice of treatments in observational studies, econometricians of the 1970s posed selection models that achieve point identification by making strong assumptions that relate realized treatment choices to treatment response. These models suffer from their own credibility problems, as discussed in Section 3.2.2.

### 3.3.1. Design-Based Inference

From the 1980s onward, some economists have argued that so-called *design-based inference* eliminates any need for selection modeling and maximizes the credibility of study of treatment response. Angrist and Pischke (2010) used the term *credibility revolution* to advocate for such analysis.

Modern advocacy of design-based inference has roots in the work of Donald Campbell and collaborators; e.g., Campbell and Stanley (1963). Campbell distinguished between the internal and external validity of studies of treatment response. A study is said to have *internal validity* if it has credible findings for the study population, whatever it may be. It has *external validity* if an invariance assumption permits credible extrapolation to a population of substantive interest. Campbell argued that studies of treatment response should be judged primarily by their internal validity and secondarily by their external validity. This perspective has been used to argue for the primacy of experimental research over observational studies, whatever the study population may be.

Campbell's doctrine of the primacy of internal validity has been extended from randomized trials to observational studies. When considering the design and analysis of observational studies of treatment response, Campbell and his collaborators recommended that researchers aim to emulate as closely as possible the conditions of an ideal randomized experiment, even if this requires focus on a study population that differs materially from the population of interest. This has led to development of various methodologies for research on so-called *quasi-experiments*.

Since the mid-1990s, the Campbell perspective has been championed by microeconomists who advocate study of a *local average treatment effect* (LATE). This is defined as the average treatment effect within the sub-population of so-called *compliers*, these being persons whose received treatments would be modified by hypothetically altering the value of a covariate called an *instrumental variable*; see Imbens and Angrist (1994). Local average treatment effects are not quantities that are relevant to decision making; see

Manski (1996, 2007), Deaton (2010), and Heckman (2010). Their study has been motivated by the fact that they are point-identified given certain assumptions that are sometimes thought credible.

Applications of instrumental variables in economics focus attention on the credibility of the key assumption that a covariate is a "valid instrument," a term that has multiple formal interpretations in the literature. Curiously, some recent research outside of economics shows less concern with the validity of instruments. Some studies advocate using numerous covariates as instruments, even if many of the potential instruments are "invalid"; see, for example, Hartford et al. (2021) and Kang et al. (2015). Many examples can be found in epidemiological research on health outcomes using data on genetic markers as potential instruments for modifiable risk factors in so-called *Mendelian randomization* studies; see Davies et al. (2018) for a review.

## 4. Will AI Solve the Missing-Data Identification Problem?

### 4.1. How Do ML Methods Cope with Missing Data?

To our knowledge, the machine learning research that underlies the current structure of AI generally copes with missing data in much the same way that empirical social scientists, statisticians, and others have long done so—deletion of incomplete cases, interpolation, imputation, weighting—relying on implicit or explicit assumptions of missingness at random. We cannot eliminate the possibility that some proprietary ML research uses other approaches that are not known to us. However, the notable developments we have found in published research mainly use imputations (aka synthetic data). Applications range from deep learning algorithms developed using simulated CT scan images to fill in obscured portions of CT scans—so-called "metal artifacts"—that arise from metal in the body (Selles et al., 2024) to generative adversarial network (GAN)-based algorithms where one algorithm generates imputed values and the adversary algorithm seeks to determine which values were imputed (Shabazian and Greco, 2023). We have found no use of weak assumptions that yield partial rather than point identification. Statisticians and social scientists have commonly assumed MAR, so it should come as no surprise that it is the standard practice in ML as well.

Some recent ML research has drawn attention to deviations from MAR. Mitra et al. (2023), for example, appear to have coined the phrase "structured missingness" (SM), which they describe as follows (p. 13): "an increasingly encountered problem…in which missing values exhibit an association or structure, either explicitly or implicitly." The SM notion seems reminiscent of missingness not at random (MNAR). They conclude with a discussion of how vast amounts of training data are being utilized, arguing (p. 20):

> "For ML methods to learn from such dynamic, heterogeneous data, and generalize robustly, they need to be designed to cope with the inevitable SM. These concerns are above and beyond issues of model degradation and bias associated with standard data cleaning processes. For this reason, we believe that there is now an urgent need to tackle SM as a topic in its own right, of central importance to the future of ML."

We agree with the urgency of taking missing data problems seriously. However, we have been struck by the apparent lack of awareness among ML researchers of the work of econometricians and statisticians on point identification of models of MNAR, much less the work on partial identification discussed in Section 3. MAR seems to be taken as the default assumption.

For example, in a "survey on missing data in machine learning," Emmanuel et al. (2021) describe three possible missing data mechanisms, missing completely at random (MCAR), MAR, and MNAR, the last of which is characterized as follows (p. 4): "Handling the missing values is usually impossible, as it depends on the unseen data." They then assert: "Many researchers, however, report that the easiest way is to complete all the missing data as MAR to some degree because MAR resides in the middle of this continuum." We see no foundation to assert a continuum from MCAR to MAR to MNAR, with MAR in the middle.

Further, we are concerned by the common belief that the availability of high-dimensional data makes an MAR assumption credible. One must ask when, if at all, MAR should be expected to hold.

### 4.2. Will Machines Do Better Than Humans?

We now return to the original question that motivated this paper, namely, the implications for research on decision making with missing data of the quest for AGI and artificial superintelligence. Based on what we understand of the current ML-based architecture of AI systems, we do not believe that machines of this type will do better than humans at developing credible assumptions to solve the identification problem created by missing data.

Our conclusion should not be interpreted as disparaging or even questioning the remarkable computational advances of AI systems, nor do we question whether advances will continue. But, while remarkable, we see these advances as continuing the long history of technological change that augments human decision making and actions. We see no reason to expect that the current trajectory of AI will surpass human intelligence, at least in the context of coping with missing data.

We are concerned that, as is consistent with the history of technology, new developments in AI will not always be beneficial. Humans may overestimate the capabilities of and inappropriately rely on their machine assistants. We highlight two potential problems.

First, as discussed above, belief in the power of additional covariates to justify MAR assumptions paired with utilization of high-dimensional data may lead practitioners to mistakenly believe that missing data is no longer a problem. Second, an important aspect of intelligence must be an ability to disregard erroneous information. Unfortunately, ML research has paid scant attention to the quality of available data, focusing instead on its quantity. This renders methods for imputation of missing values even more problematic.

Although we believe that machines are not now on a path to credibly shrink the state space for missing data, we do not dismiss the possibility that AI will someday develop a superior approach to decision making under uncertainty—some might say "a new paradigm"—that we cannot currently imagine and that perhaps no human would imagine in the foreseeable future.

We emphasized in Section 2 that various criteria for reasonable decision making under uncertainty have been developed, among which there is no consensus and no clear way to choose among them. It would be foolhardy for us to dogmatically assert that a different form of intelligence will never find a clear choice that we cannot currently comprehend, much as humans in the 19$^{th}$ century could not yet comprehend the theory of relativity and quantum mechanics. Nevertheless, while acknowledging our limited capabilities as humans with bounded rationality, as well as a lack of knowledge about ongoing proprietary research across the globe, we firmly believe that the current ML-based path of AI will not get us there.

*The Promise and Limits of Deep Learning Algorithms*

To illustrate why we are skeptical, we point to current research on and beliefs about deep learning algorithms. As researchers with considerable experience developing and applying nonparametric statistical methods, we are well aware of the implications of the *curse of dimensionality* in conditional prediction; that is, the increasing difficulty of prediction as the dimension of the covariate vector increases. It appears, however, that some ML proponents and practitioners believe that their methods have broken the curse, enabling essentially assumption-free and accurate predictive algorithms with high-dimensional data.

An early and influential proponent of this line of thinking was the statistician and early ML researcher Leo Breiman, who wrote of the issue in his 2001 *Statistical Science* article entitled "Statistical Modeling: The Two Cultures." With regard to assumptions, he asserted the following about what would subsequently become the dominant approach to building predictive algorithms (Breiman, 2001, p. 205): "The one assumption made in the theory is that the data is drawn i.i.d. from an unknown multivariate distribution." He went on to question the conventional wisdom that "high dimensionality is dangerous," countering that "recent work has shown that dimensionality can be a blessing." He continued (p. 208):

> "Reducing dimensionality reduces the amount of information available for prediction. The more predictor variables, the more information. There is also information in various combinations of the predictor variables. Let's try going in the opposite direction: Instead of reducing dimensionality, increase it by adding many functions of the predictor variables."

This advice, which has helped shape ML research over the past 25 years, reminds us of the common advice to condition on more and more covariates to justify MAR assumptions for coping with missing data.

Some recent ML research acknowledges that the curse of dimensionality has not, in fact, been broken. (This is heartening although, given that the curse is a well-understood mathematical property, it should never have been in question.) We point here to Poggio and Fraser (2024) who, studying the compositional sparsity that underlies deep learning algorithms, write (p. 438):

> "Compositional sparsity, or the property that a compositional function have *[sic]* 'few' constituent functions, each depending on only a small subset of inputs, is a key principle underlying successful learning architectures."

Recognition of the centrality of a sparsity assumption to justify deep learning methods, while seemingly underappreciated in the ML research community, is not new. In his discussion of Schmidt-Hieber (2020) concerning "nonparametric regression using deep neural networks," Shamir (2020) concluded (p. 1912): "Essentially, we have replaced a 'curse of dimensionality' effect with a 'curse of sparsity'."

We conclude that, although some may believe that the current architecture for ML-based AI systems with access to vast troves of training data will be able to develop finely tuned predictive models while credibly assuming MAR holds, arguments of the type made by Breiman and Rubin do not hold up. There is no magic in applying MAR assumptions to missing data by conditioning on numerous covariates.

## 5. Broader Implications

To assess the implications of ongoing research in AI for methodological research on decision making with missing data, we have focused attention on two leading cases: missing outcome data in surveys and the unobservability of counterfactual outcomes. In each case, we have sought to clarify both (i) the conditions for application of credible assumptions that shrink the state space and improve decision making and (ii) why we do not believe that the current dominant ML-based AI architecture will enable machines to do better than humans at developing credible assumptions to solve these identification problems, which we would take as evidence of superintelligence.

There are many important missing data problems beyond the two that we discussed. Perhaps the most recognizable is *extrapolation* (aka external validity), whereby a decision maker seeks to apply findings from one population to a different population. Successfully addressing this identification problem requires developing credible assumptions that connect the latter distribution to the former. Bounded-variation assumptions may shrink the state space and stronger assumptions may yield point identification, but there is no magic bullet. The Law of Decreasing Credibility still applies both to humans and to machines.

Econometrics has long recognized other identification problems, including those that arise from dimensions of data quality beyond missing data. The consequences of white-noise measurement error have been studied since the 1930s (Frisch, 1934). Statisticians have studied the consequences of contaminated and corrupted data from the perspective of robust statistical analysis (Huber, 1981), with later interpretation as an identification problem by econometricians (Horowitz and Manski, 1995). In each case, the identification problems that arise require developing credible assumptions to shrink the state space. We do not see how, on its current path, ML-based AI will surpass humans in this regard. As noted above, ML research has paid scant attention to the quality of available data, focusing instead on its quantity.

A different type of identification problem in all of the sciences regularly arises from model uncertainty. Modeling climate change is a prominent example. Climate scientists have been well aware that alternative climate models yield a wide range of forecasts of the trajectory of future global warming. Manski et al. (2021) frame climate modeling as a problem of partial identification and propose an approach that integrates leading models using the minimax-regret decision criterion we have discussed here.

A fundamental problem facing climate modelling is that there exists only one climate history from which to collect data. Without the possibility of repeated experimentation, it is not surprising that multiple climate models will fit the available data reasonably well. We do not know how much AI will help us make progress in understanding climate change. We are firm, however, in our belief that Occam's razor need not apply to climate modelling; that is, the simplest climate model need not be the most credible. We also conjecture that black box, big data methods will not be able to solve the identification problem in climate modeling. We expect that any solution will require incorporating knowledge and credible assumptions about the structure of the problem.